\documentclass{aa}
\newcommand{\um}{\ensuremath{\mu}m\,}
\newcommand{\mh}{H$_2$\,}
\newcommand{\kms}{km\,s$^{-1}$\,}
\renewcommand{\deg}{\ensuremath{^\circ}}
\usepackage{txfonts}
\usepackage{natbib}
\usepackage{graphicx}
\usepackage{geometry}
\begin{document}
   \title{Investigating the transport of angular momentum from young stellar objects}

   \subtitle{Do \mh jets from class I YSOs rotate?}

   \author{A. Chrysostomou
          \inst{1}
			F. Bacciotti
			\inst{2}
			B. Nisini
			\inst{3}
			T.P. Ray
			\inst{4}
			J. Eisl\"{o}ffel
			\inst{5}
			C.J. Davis
			\inst{6}
			M. Takami
			\inst{7}
          }

   \offprints{A. Chrysostomou}

   \institute{$^1$ Centre for Astrophysics Research, Science \& Technology Research Institute, University of Hertfordshire, Hatfield, HERTS AL10 9AB, UK
              \email{a.chrysostomou@herts.ac.uk} \\
            $^2$ INAF-Osservatorio Astrofisico di Arcetri, Largo E. Fermi 5, I-50125, Florence, Italy \\
			$^3$ INAF-Osservatorio Astronomico di Roma, Via di Frascati 33, I-00040 Monte Porzio Catone, Italy \\
			$^4$ School of Cosmic Physics, Dublin Institute for Advanced Studies, 5 Merrion Square, Dublin 2, Ireland \\
			$^5$ Th\"{u}ringer Landessternwarte, Sternwarte 5, D-07778 Tautenberg, Germany \\
			$^6$ Joint Astronomy Centre, 660 N. A'ohoku Place, Hilo, Hawaii 96720, USA \\
			$^7$ Subaru Telescopes, 650 N. A'ohoku Place, Hilo, Hawaii 96720, USA 
              }

   \date{Received ---; accepted ---}

 
  \abstract
   {}
   {In this pilot study, we examine molecular jets from the embedded Class I sources, HH\,26 and HH\,72, to search, for the first time, for kinematic signatures of jet rotation from young embedded sources.}
   {High-resolution long-slit spectroscopy of the \mh 1-0\,S(1) transition was obtained using VLT/ISAAC. The slit was placed perpendicular to the flow direction about 2$\arcsec$ from the sources. Position-velocity (PV) diagrams are constructed and intensity-weighted radial velocities \textit{transverse} to the jet flow are measured. }
   {Mean intensity-weighted velocities vary between $v_{\rm LSR} \sim$ -90 and -65\,\kms for HH\,26, and -60 and -10\,\kms for HH\,72; maxima occur close to the intensity peak and decrease toward the jet borders. Velocity dispersions are $\sim 45$ and $\sim 80$\,\kms for HH\,26 and HH\,72, respectively, with gas motions as fast as -100\,\kms present. Asymmetric PV diagrams are seen for both objects, which a simple empirical model of a cylindrical jet section shows could in principle be reproduced by jet rotation alone. Assuming magneto-centrifugal launching, the observed HH\,26 flow may originate at a disk radius of 2-4\,AU from the star with the toroidal component of the magnetic field dominant at the observed location, in agreement with magnetic collimation models. We estimate that the kinetic angular momentum transported by the HH\,26 jet is $\sim 2 \times10^{-5}$\,M$_\odot$\,yr$^{-1}$\,AU \kms. This value (a lower limit to the total angular momentum transported by the flow) already amounts to 70\% of the angular momentum that has to be extracted from the disk for the accretion to proceed at the observed rate.
   }
   {These results of this pilot study suggest that jet rotation may also be present at early evolutionary phases and support the hypothesis that they carry away excess angular momentum, thus allowing the central protostar to increase its mass.}


	\authorrunning{Chrysostomou et al.}

	\titlerunning{Transport of angular momentum from YSO jets}

   \maketitle
%

\section{Introduction}

The appearance of jets from young stars and active galactic 
nuclei is arguably one of the most enigmatic and strikingly 
beautiful phenomena in astrophysics. Yet we understand 
very little about the detailed physical processes which contrive 
to produce these jets and still debate the general mechanisms 
responsible for them \cite[for the most recent reviews see][]{Bally07,Ray06}. The inflow and outflow of material is believed to be mediated by the interaction of magnetic and centrifugal forces in the close environment of a newborn star. In particular, a magnetised accretion disk channels material towards the protostar (on which it will finally be accreted via coronal loops) and provides the necessary magnetic forces to launch material off the disk surface into a jet/outflow. One fundamental problem to which 
jets are believed to provide a solution, is the necessary removal of excess angular momentum from the circumstellar disk. This 
is required to permit material to fall to low angular momentum orbits and thus be finally accreted onto the protostar, thereby increasing its mass up to its final value. In addition, jets have the important 
role of dispersing the infalling material from the surrounding 
circumstellar envelope and hence influence the final mass of 
the central protostar. 

Recent observations give evidence of a small difference between the radial velocities of lines emitted from two opposed sides of a jet (with respect to the symmetry axis). This was interpreted in the context of gas rotation. The first tentative 
results came from observations of the HH\,212 outflow by \cite{Davis00}. Velocity shifts of a few \kms were found for the \mh lines across the breadth of the jet. However, the rather coarse spatial and spectral resolution of the data prevented any definitive conclusions, especially given that the region of the jet investigated in that study is located at not less than $10^4$\,AU from the source, too far from the acceleration zone (typically a few AU above the disk) to exclude interaction with the jet environment. Independently, using the 
STIS spectrometer on the \textit{Hubble Space Telescope}, first \cite{Bacciotti02} and then \cite{Coffey04,Coffey06} and \cite{Woitas05} presented better evidence for jet rotation from several small-scale jets emanating from classical T-Tauri stars, including DG Tau, RW Aur, Th 28 and CW Tau. These high angular resolution optical and near-ultraviolet observations, with slits placed parallel and perpendicular to the flow axis, are from a region much closer to the acceleration engine ($\sim 20-100$\,AU). The derived rotational motions appear to be in agreement with theoretical predictions of models developed for magneto-centrifugally launched jets \citep{Anderson03,Pesenti04}, supporting the idea that jets do transport angular momentum. These observations, however, only consider jets from evolved sources (Class II YSOs) in the last phases of their pre-main sequence evolution.

If these signatures are evidence of jet rotation, then to establish the importance of this mechanism for star formation we need to determine if such features are also seen at earlier evolutionary epochs, such as Class I YSOs. Current theories would predict this since inflow/outflow is present as a disk forms around the protostar \citep[see][]{Pudritz06,Shang06}. \cite{Davis02} demonstrated that a small number of Class I YSOs harbour small scale \mh jets, so-called molecular hydrogen emission line regions (MHEL -- see \cite{Davis01}) extending some $2-3\arcsec$ from the central infrared source. These sources give us the opportunity to conduct ground-based searches for and comparisons of the kinematics of jet rotation between evolved T-Tauri and young Class I YSOs.

Here we report on a pilot study, conducted on the VLT, which suggest that jets from Class I YSOs are rotating as they emerge from the central engine (\S3), and present a simple kinematic model which, at least for HH\,26, supports this suggestion along with analysis assuming that the jet is launched magneto-centrifugally (\S4). Finally, we comment on the relevance of these findings in the context of jet rotation studies (\S5).


\section{Observations}

\begin{table*}
\begin{minipage}[t]{\textwidth}
\caption{Observing Log and source parameters.}             
\label{log}      
\centering                       
\renewcommand{\footnoterule}{}  
\begin{tabular}{c c c c c c c}        
UT date\footnote{All UT dates in this paper are presented as YYYY-MM-DD.} & Object & RA (J2000) & Dec (J2000) & Slit PA (\deg) & Distance (pc) & Integration time\\    
\hline                        
   2004-02-07 & HH\,26 & 05 46 5.1 & -00 14 17 & 130 & 400\footnote{\cite{Anthony-Twarog82}} & $8\times300$ sec \\      
   2004-03-02 & HH\,72 & 07 20 10.3 & -24 02 24 & 0  & 1500\footnote{\cite{Davis01}} & " \\
   2004-03-05 & HH\,26 &      &  & 130 & & " \\
   2004-03-08 & HH\,72 &     &  & 0 &  & " \\
\hline                                   
\end{tabular}
\end{minipage}
\end{table*}

\subsection{Source selection}

The jets from HH\,26 and HH\,72 were selected for this study. In selecting suitable Class I YSOs to search for jet rotation we are limited to using tracers longward of the optical. This is because sources this young are always found embedded within molecular clouds which also makes the use of adaptive optics very difficult in the absence of a nearby optical guide star. The 2.12\,\um v=1-0\,S(1) transition of \mh is an excellent tracer of the dynamics of outflows and jets in embedded sources \citep{Bally07}. At present, the work of \cite{Davis01,Davis02} represents the best source of YSOs with \mh jets and two candidates from this work were selected to conduct our pilot study. HH\,26 was chosen because it is relatively nearby and bright (in terms of available MHEL sources). Furthermore, it is a well studied object with published ancillary data to support our analyses. HH\,72 is less well studied and is much further away, but represents an unique opportunity to search for rotation in the jet from a more massive YSO. 

Our objectives with this exploratory study is to establish the presence or absence of evidence for jet rotation in these young sources before embarking on a more extensive study.

\subsection{Observations \& data reduction}

The data were obtained using the ISAAC instrument on \textit{VLT-ANTU} over several nights in February and March 2004 (see observing log in Table \ref{log} for details).

\begin{figure*}
\centering
\includegraphics[width=7.5cm]{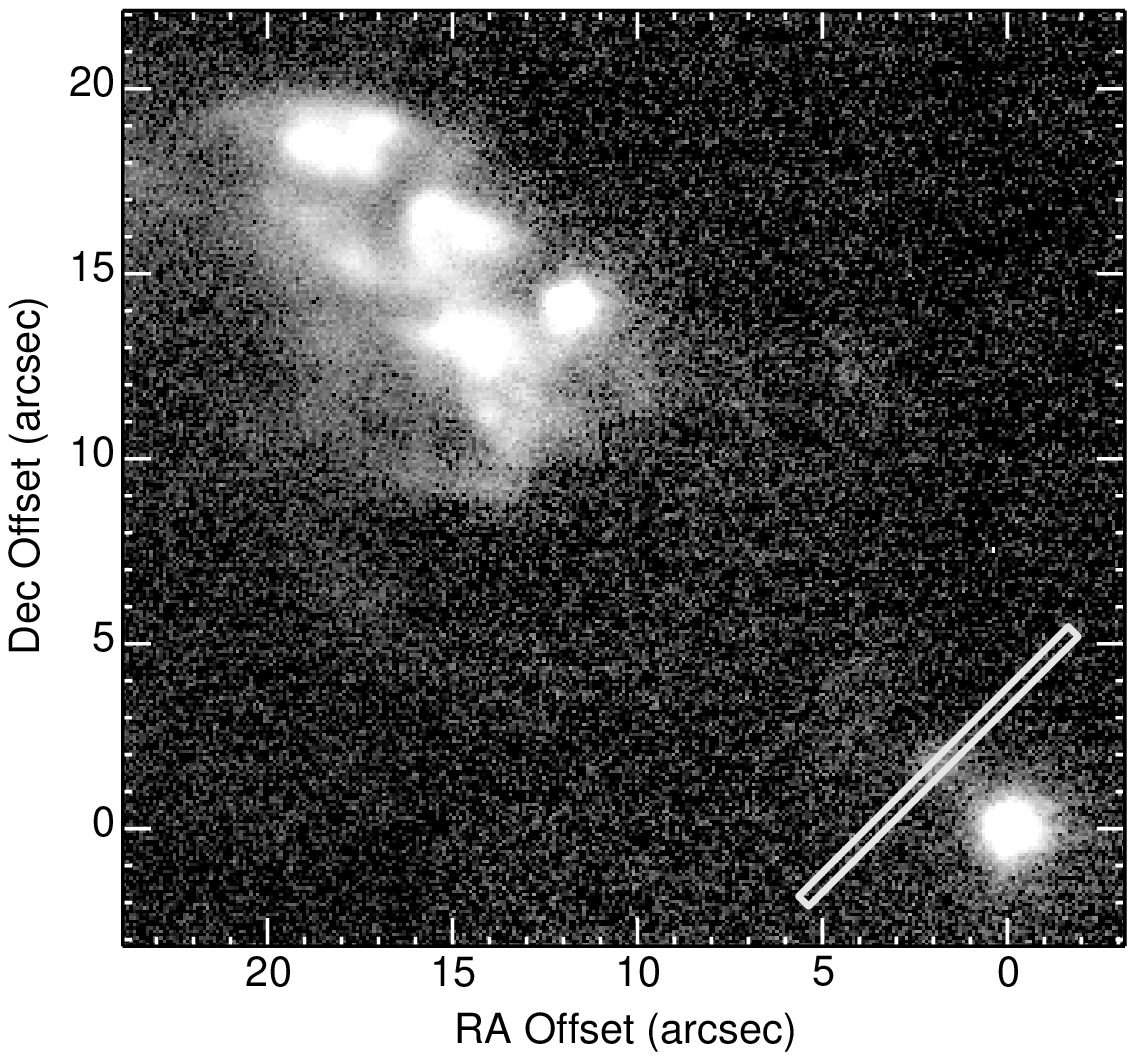}
\includegraphics[width=10.3cm]{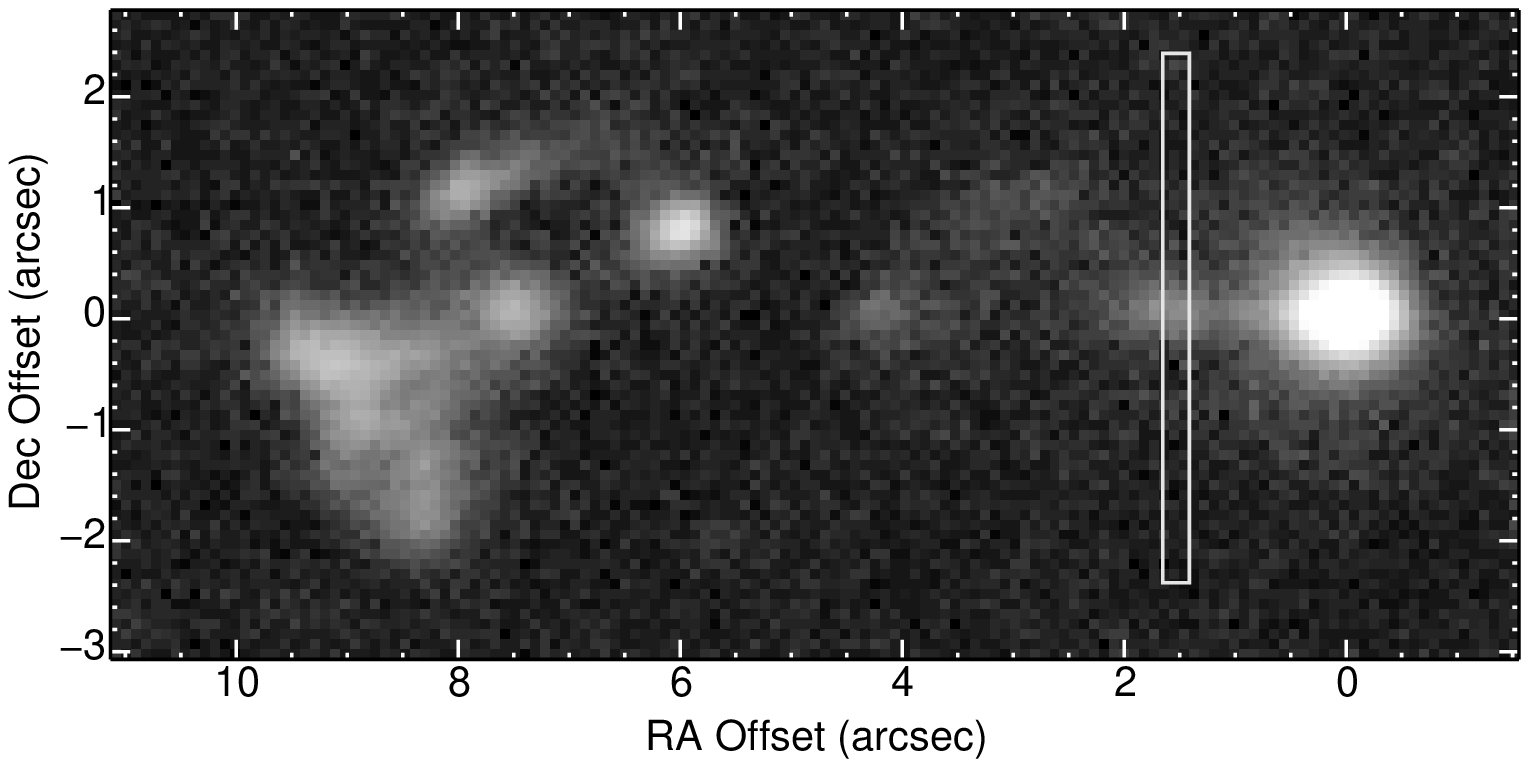}
\caption{\mh\,v=1-0\,S(1) images of the HH\,26 (left) and HH\,72 (right) outflows \cite[adapted from][]{Davis02}. In each panel part of the larger scale flow is seen together with the YSO driving the outflow (from which the coordinates are referenced). The thin white boxes are representations of the slit placement and their position angles. The observations were targeted to the \mh~jet close to the source with the slit placed perpendicular to the flow axis (see Section 2 for details).}
\label{slits}
\end{figure*}

ISAAC was configured using the SWS1-MR, medium resolution spectroscopy mode and centred on the v=1-0 S(1) transition of \mh at $\lambda = 2.1218$\,\um. A spectral resolution of R=8900 together with the $0.3\arcsec$ slit provided a velocity resolution of 33.7\,\kms (or $\sim$ 17\,\kms per pixel). The instrument optics of $0.148\arcsec$ per pixel ensured good sampling of the emission structures ($\sim 2-3\arcsec$). Each exposure through the long-slit spectrograph was 300 seconds.
In some previous studies, slits were placed parallel to the flow and stepped across, building up a three-dimensional view of the jet \citep[e.g.][]{Davis00,Bacciotti02,Woitas05}. However, the unavoidable uneven illumination of the slits introduced spurious velocity shifts which had to be calculated and subtracted from each spectrum. In this work, the slit is placed perpendicular to the flow axis -- at a position angle of 130\deg\, and 2\,\arcsec\, from the source of HH\,26, and at a position angle of 0\deg\, and 1.5\,\arcsec\, from the star driving the HH\,72 flow (see Figure \ref{slits}) -- thus sampling the velocity field across the flow in one exposure \citep[see also][who also observed in this manner]{Coffey04,Coffey06}. 

A standard $ABBA$ observing sequence was adopted to remove the sky background with the $A$ and $B$ telescope beam positions separated by $60\arcsec$ along the slit, sufficient to avoid any overlap of \mh emission and keep the source on the array at all times. The data were reduced using standard techniques and \textsc{starlink} software packages. Image straightening and wavelength calibration was performed using OH sky lines \citep[using the line list of][]{ohlines}. The final wavelength calibration was accurate to $\sim 0.1$\,\AA~ (rms dispersion of $\sim 1.3$\,\kms). 
The velocity scale was calculated relative to the vacuum wavelength of the \mh\,v=1-0\,S(1) transition, given as 2.121833\,\um by \citet{BvD87}. Velocities have been corrected for the velocity of the parent clouds\footnote{ $-33$ \kms and $+10$ \kms for HH\,72 and HH\,26 respectively.} and the Earth's relative motion on the nights observed. Position-velocity (PV) diagrams are constructed after aligning and combining the $A$ and $B$ beams. The seeing varied between $0.5\arcsec$ and $1.0\arcsec$.

\section{Results}

\subsection{PV diagrams}

\begin{figure*}
\begin{minipage}[]{\textwidth}
\centering
\includegraphics[height=7cm]{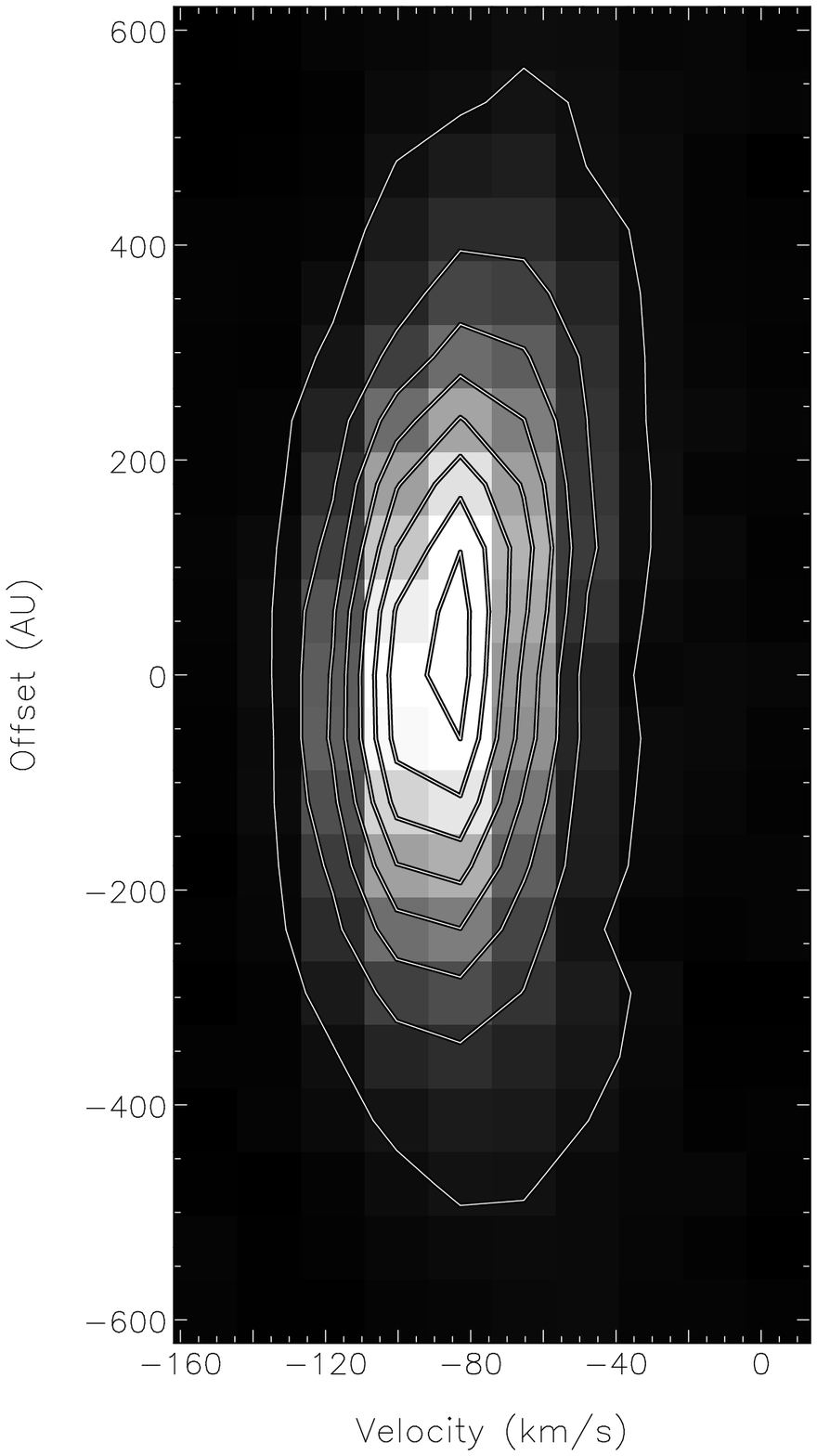}  
\includegraphics[height=7cm]{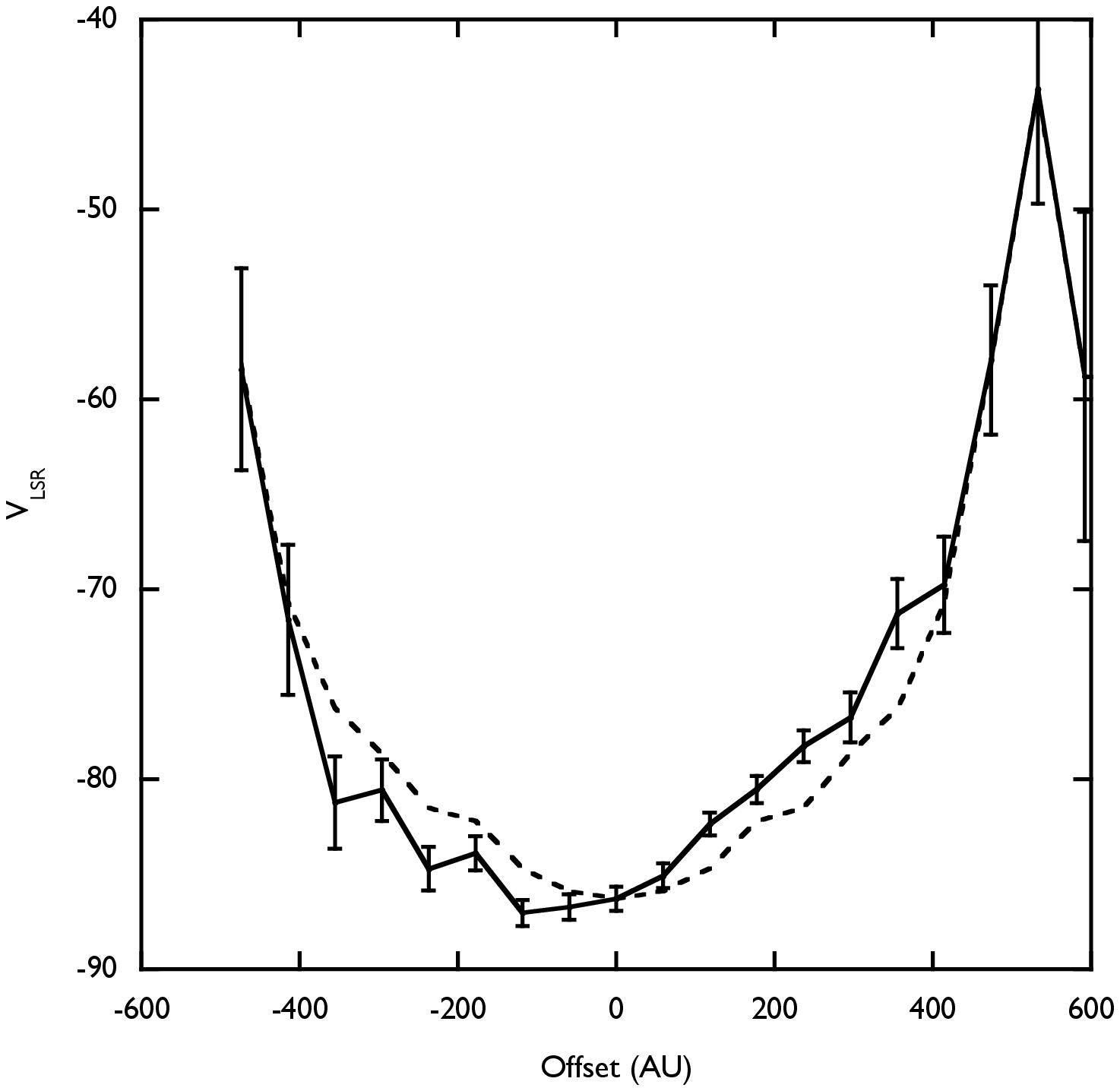}
\vspace{6mm}
\end{minipage}
\begin{minipage}[]{\textwidth}
\centering
\includegraphics[height=7cm]{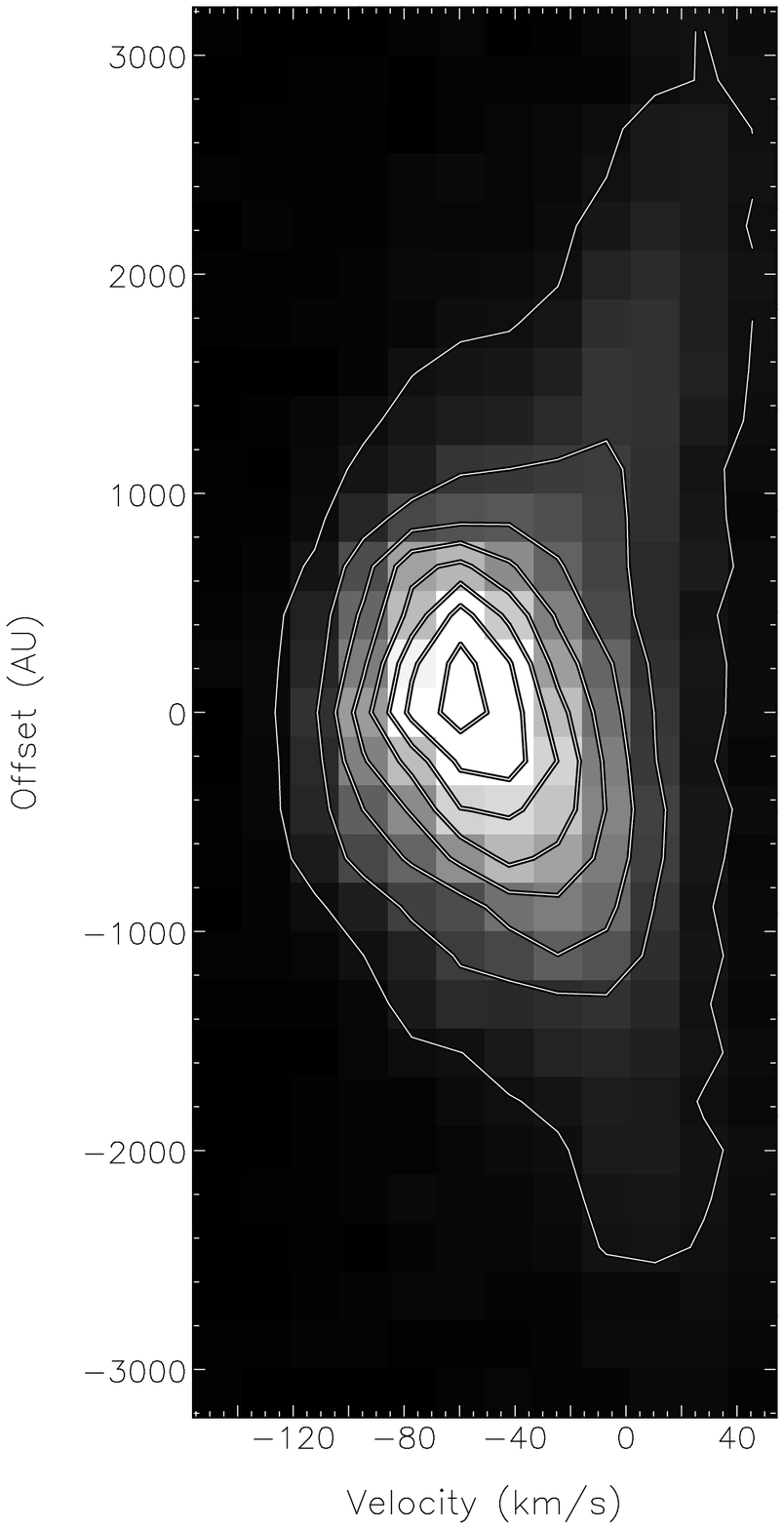}
\includegraphics[height=7cm]{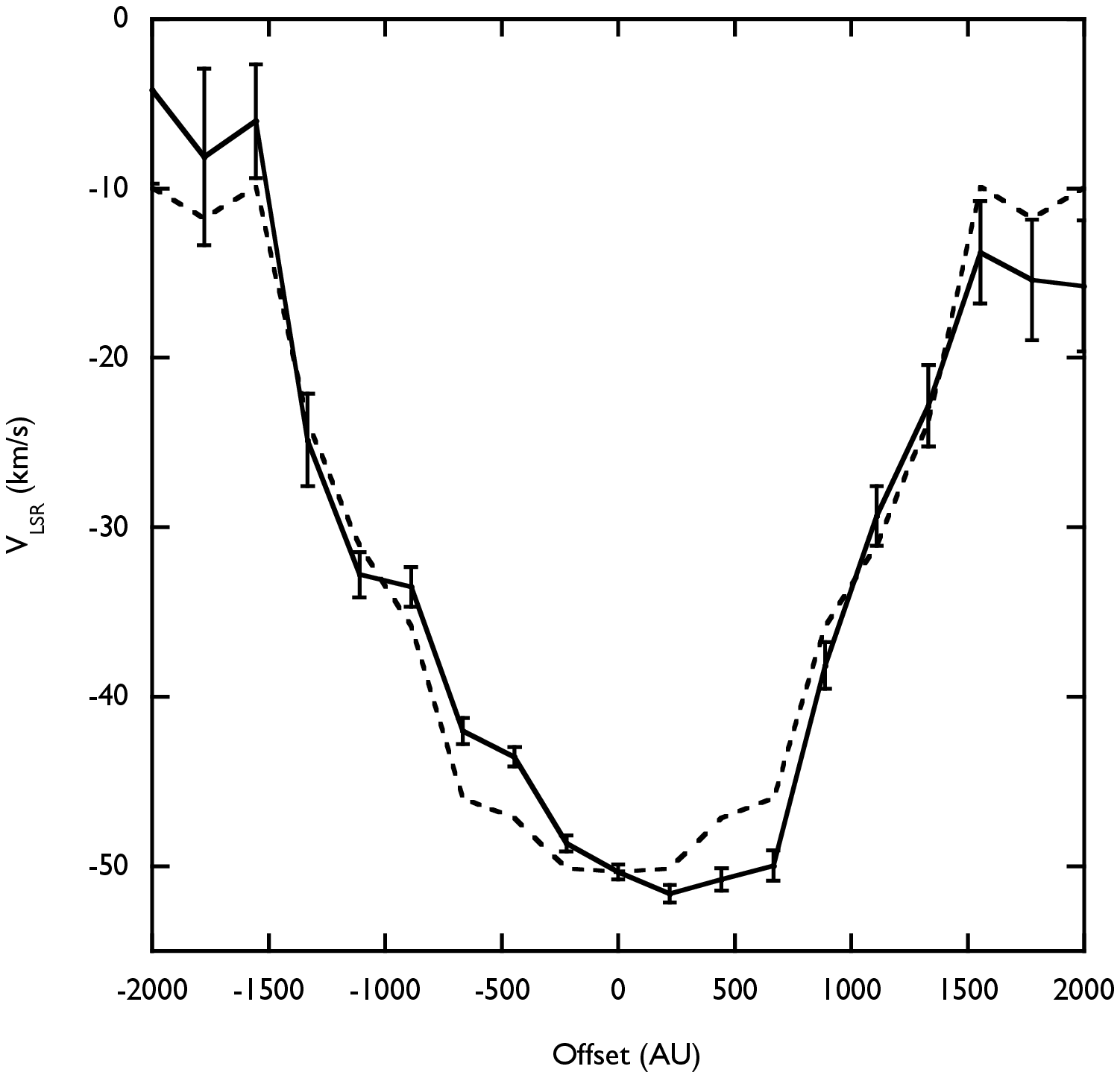}
\end{minipage}
\caption{PV and radial velocity diagrams for HH\,26 (top) and for HH\,72 (bottom) measured at a distance of $2\arcsec$ and $1.5\arcsec$ from the source, respectively. The left hand panels show the PV diagrams and the right hand panel shows the mean intensity-weighted radial velocity measurements (solid line); the dashed line shows the symmetry profile (the average of velocities either side of the axis) and approximates the profile in the absence of rotation. Offsets (in AU) are measured along the slit and relative to the pixel with highest flux. At the assumed distances to each object (see Table \ref{log}) 1000 AU corresponds to $2.6\arcsec$ in HH\,26 and $0.7\arcsec$ in HH\,72. }
\label{pvdiag}
\end{figure*}

In Figure \ref{pvdiag} we present the position-velocity (PV) diagrams extracted from the data for HH\,26 and for HH\,72. The slit offset is centred on the photocentre (brightest pixel) and is explicitly assumed to define the outflow axis. The diagrams show that the jet velocities are quite high, of the order of $-60$ to $-100$~\kms, while the velocity dispersion for HH\,26 is narrower ($\sim 45$\,\kms) than that for HH\,72 ($\sim 80$\,\kms). 

The mean intensity-weighted radial velocity (measured on the axis) for HH\,72 is $-56.3 \pm 0.3$~\kms, consistent with previous \mh echelle spectroscopy for this object \citep{Davis01}. Although observed at lower spatial resolution, so that the particular component presented here was not properly resolved\footnote{The MHEL jet in both HH\,26 and HH\,72 were later identified using Fabry-Perot imaging \citep{Davis02}.}, \cite{Davis01} did detect emission in their slit up to velocities of $\sim -165$~\kms. 
On source, they report a complex of emission features with components at $+13$, $-40$ and $-129$\,\kms. As the slit for these data was placed $\sim 1.5\arcsec$ from the source, we are probably detecting the same component responsible for the -40\,\kms peak. If so, then the jet either accelerates away from the source or was faster in the past.

For HH\,26, the peak on-axis radial velocity is $-87.5 \pm 0.4$~\kms. No published \mh spectroscopy towards the HH\,26 MHEL is known. \cite{Davis00} present echelle spectroscopy of the \mh emission measured in the HH\,26A and C outflow lobes (HH\,26A is the lobe prominent in Figure \ref{slits} while HH\,26C is further to the north-east and not shown here) with the slits placed \textit{along} the outflow axis. Their data showed emission up to velocities of $\sim -50$\,\kms in HH\,26A and $\sim -90$\,\kms for HH\,26C. 

For a slit placed perpendicular to the flow direction, a signature of rotation is apparent as asymmetric contour lines in a PV diagram. Assuming axisymmetry, material moving towards and away from us at a given longitudinal radius is slightly more blue- and red-shifted with respect to material moving transversely to the line of sight. These differences are more evident at large distances from the axis than closer in \citep[e.g. see][]{Coffey04} because of projection and convolution effects \citep{Pesenti04}.

Asymmetry is present in both PV diagrams shown in Figure \ref{pvdiag}, the effect more pronounced in HH\,72 than in HH\,26. The figures also show that this feature is not only different in magnitude in the two objects but also of opposite sense (the direction of the putative rotation determining which way the contour lines slope). Importantly, this indicates that it is unlikely that our results are due to an unforeseen or unaccounted for systematic error.

\subsection{Radial velocity measurements}

Figure \ref{pvdiag} also shows the mean intensity-weighted radial velocities measured for HH\,26 and HH\,72 along the slit and displayed as a function of distance from the axis. This velocity was calculated for each row along the slit by determining,

\begin{equation}
\langle v \rangle = \frac{\sum_i\; I_i v_i}{\sum_i\; I_i}
\end{equation}

\noindent where the summations are made over the whole spectrum. To determine the mean value and its uncertainty, each pixel intensity, $I_{i}$, has its value randomly adjusted by either adding or subtracting its error, or not at all. This is repeated 1000 times for each pixel row thus establishing a distribution for $\langle v \rangle$. 

What is immediately striking in Figure \ref{pvdiag} is the decrease in absolute velocity away from the axis of the flow. This has already been observed in a number of T-Tauri jets at angular resolutions high enough to resolve the flow width -- of the order of 100 AU at a similar distance from the source \citep{Bacciotti00,Woitas05,Coffey04,Coffey06}. A decrease in poloidal velocity is expected from magneto-centrifugal launch theories, the Keplerian speeds in the launching zones of the disk wind naturally providing a wide range of poloidal jet velocities across the beam \citep{Ferreira06}. In principle, it is possible that the outer portions of the flow are being slowed by interaction with the environment, however, we doubt this possibility given:  (a) the jet is hypersonic and hence is not expected to entrain much ambient material particularly close to the source \citep{Ray00}, (b) there is no significant decrease in average poloidal velocity with distance from the source \citep[e.g.][who, for HH\,26, actually report an acceleration at the position of our observation]{Calzoletti08}, and (c) recent mass flux estimates along jets \citep[e.g.][]{Podio06} suggest very little if any entrainment.

The dashed lines in Figure \ref{pvdiag} are the symmetry profiles which would be obtained for unperturbed, non-rotating flows. It is computed from $v_{\textrm{sym}} = (v_+ + v_-)/2 $, where $v_+$ and $v_-$ represent the velocities measured at positions either side of and equidistant from the axis. 

The radial velocity data appear systematically displaced from the symmetry profile, with the direction of the displacement being different/opposite for the two objects. Such a displacement would be observed if the gas were rotating and, if so, the different directions indicate that the \textit{sense} of rotation is different for the two objects. 

It is worth noting that this signature could also be generated by asymmetric bow shocks, with the \mh streaming down the bow wings at different rates. However, within the present resolution limits we see no evidence of bow shocks but rather a linear jet-like feature (see Fig. \ref{slits} and \cite{Davis02}). If the \mh emission is tracing material shocked and entrained by the jet rather than the jet itself, then in the absence of alternative solutions, the evidence suggests that it was entrained by a rotating jet. Such scenarios need to be considered and properly modelled, but are beyond the scope of this paper. In what follows we consider a simple model in which the observed PV diagrams are produced by a rotating jet.


\section{Modelling}

In this section we use analytical models designed to better understand the relation between a rotating structure and the resultant PV diagram. Moreover, assuming that these structures are indeed rotating and using very general conservation laws for magneto-centrifugally launched jets, we derive some physical parameters from the observed features.

\subsection{PV diagram model}

\begin{figure}
\includegraphics[width=\columnwidth]{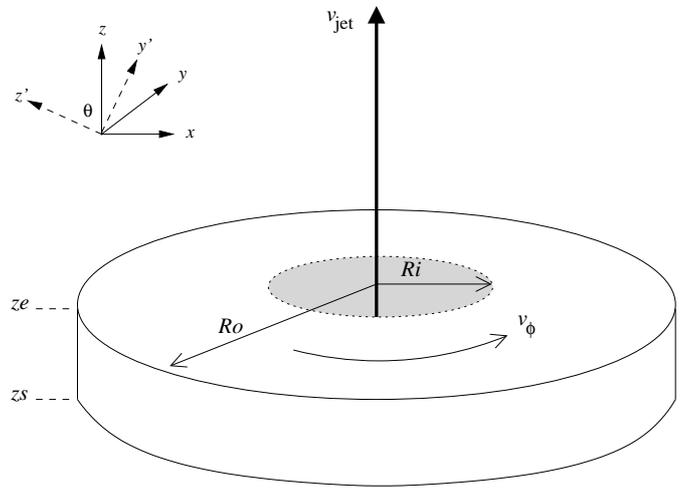}
\caption{Schematic of the structure used to model a rotating  jet beam. The coordinate scheme, whose origin is centred on $r = 0$ is also depicted. The $x$ ordinate corresponds to the impact parameter and $z$ is the height or thickness of the object projected onto the plane of the sky. The \mh emitting portion of the jet has inner and outer radii truncated at $Ri$ and $Ro$, respectively, and is rotating with a velocity $v_\phi$. The transformation following an inclination of $\theta$ about the $x$ axis (towards the line-of-sight) is shown.}
\label{schem}
\end{figure}

Figure \ref{schem} is a schematic representation of an empirical model used to describe the section of a rotating YSO jet that emits the radiation collected by our slits. The jet section is represented as a geometrically thick disk (thickness $= ze - zs$) and is inclined towards our line of sight at some angle, $\theta$. The jet section has an external radius $Ro$ and is truncated internally at a radius $Ri$ -- inner axial holes are predicted in most models of X-winds and disk winds \cite[e.g. see][]{Pudritz06,Shang06}. The density distribution for the jet section is assumed to be a function of the radius, $n(r)$, first increasing to a maximum and then decreasing smoothly outwards on a distance scale $a$. Such a distribution can be described by the function:

\begin{equation}
n(r) \propto r^\alpha \exp (-r/a)
\label{dens}
\end{equation}

The material in the jet section possesses an axial velocity which we take to be the jet velocity, $v_{\rm jet}(r)$, and is rotating at a rate $v_\phi(r)$. The radial dependence of both these velocity components are expressed as power-laws,

\begin{eqnarray}
v_\phi(r) &=& v_{\phi,0} \left( \frac{r}{Ro} \right)^{-\beta} \\
v_{\rm jet} (r) &=& v_{\rm jet,0} \left( \frac{r}{Ri} \right)^{-\gamma}
\label{vels}
\end{eqnarray}

Note that, for convenience, the azimuthal velocity is normalised to the outer edge of the structure, while the  jet velocity is determined relative to a central velocity defined at the inner truncation radius. With this simple formalism, a separate radial component $v_r(r)$ may also be added to reflect either expansion or contraction of the gas from or toward the axis. The velocity along the line of sight and at each point $(x,y)$ of the plane (ie. edge-on) of the jet section is then given by,

\begin{equation}
v(r) = v_\phi(r) \left( \frac{x}{r} \right) - v_r(r) \left( \frac{y}{r} \right)
\label{vr}
\end{equation}

\noindent where $x$ and $y$ are coordinates on a Cartesian grid with the origin coincident with $r = 0$. Allowing the structure to incline relative to the plane of the sky requires that the coordinate grid is transformed about the origin,

\begin{eqnarray}
x^\prime &=& x \nonumber \\
y^\prime &=& (y \cos \theta) + (z \sin \theta)  \\
z^\prime &=& (z \cos \theta) - (y \sin \theta) \nonumber
\end{eqnarray}

\noindent Each velocity value at position ($x$, $y$, $z$) calculated by equation \ref{vr} is then assigned to position ($x^{\prime}$, $y^{\prime}$, $z^{\prime}$).

The velocity component in the inclined plane viewed along the line-of-sight thus becomes,

\begin{equation}
v(r,\theta) = v(r) \cos \theta - v_{\rm jet} (r) \sin \theta
\end{equation}

We calculate the \mh emission from the jet section entering our slit by assuming that the emitting gas is optically thin and dust free. The synthetic image thus produced is then dispersed in wavelength according to the instrumental characteristics of the spectrograph. In this way, we obtain a synthetic PV diagram which we can directly compare to our observations.

\subsubsection{Results}

Models whose jet velocity is constant across the jet do not produce results which are representative of the data, regardless of the behaviour of the angular velocity component. Better representation is achieved if the poloidal jet velocity is allowed to decrease as a function of distance away from the axis. This alone shows that the jet velocities observed cannot be constant across the structure but must decrease with radius.

\begin{figure*}
\centering
\includegraphics[width=8.5cm]{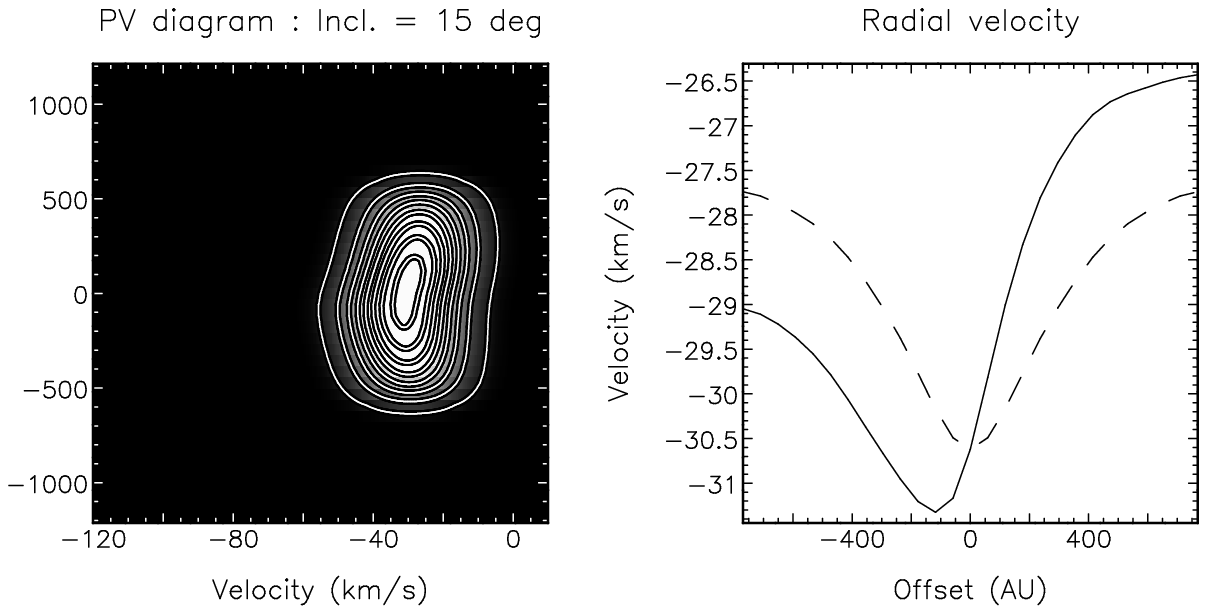}
\hspace{5mm}
\includegraphics[width=8.5cm]{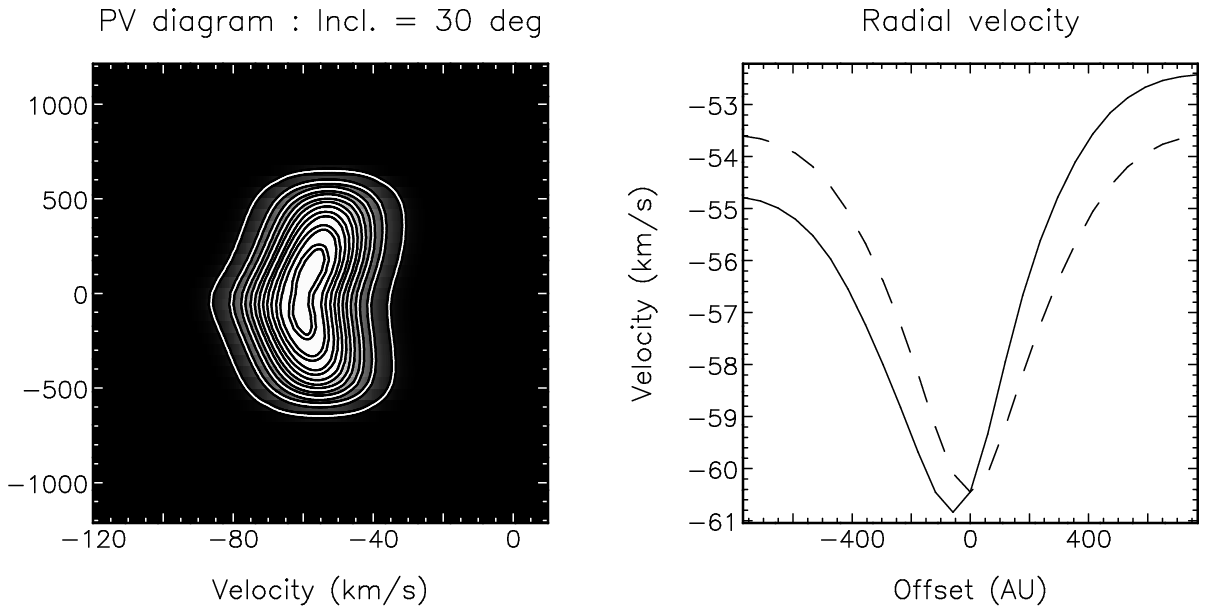}
\includegraphics[width=8.5cm]{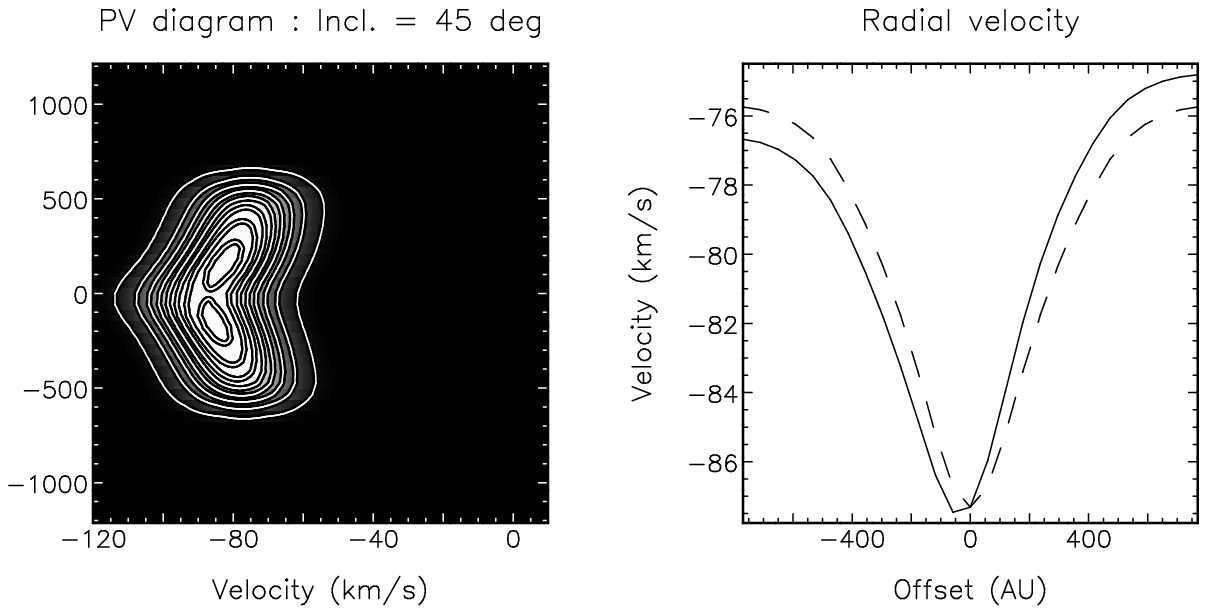}
\hspace{5mm}
\includegraphics[width=8.5cm]{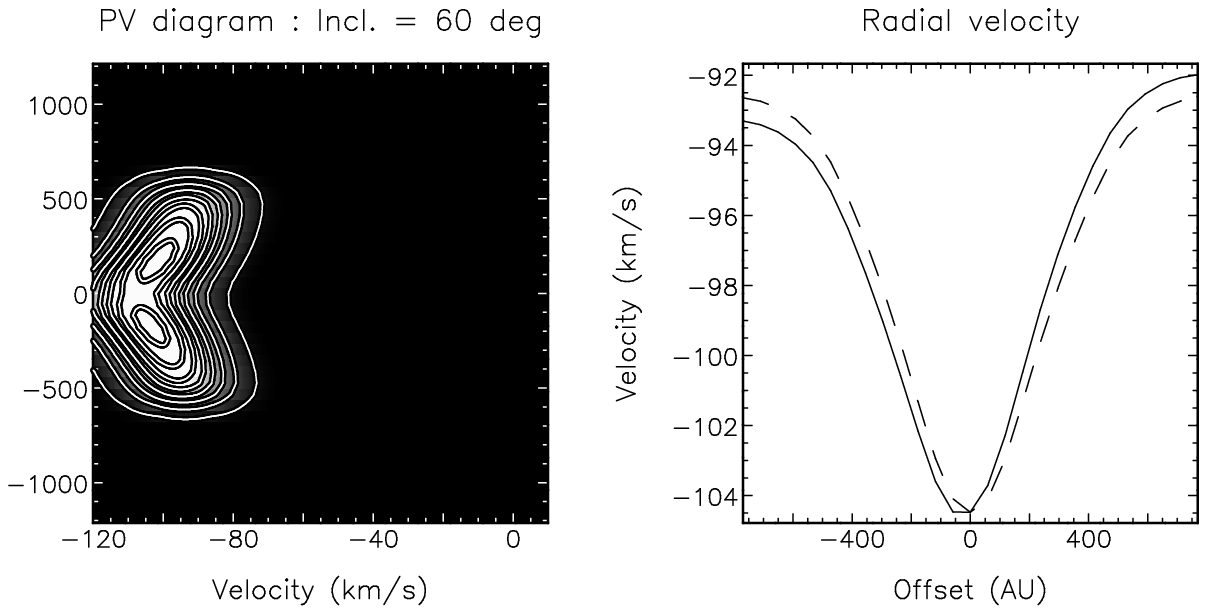}
\caption{PV diagram models. The panels show PV diagrams along with the associated positions of intensity-weighted radial velocities (solid line) compared to a symmetric velocity profile (dashed line), i.e. with no rotation, beside them. The inclination of the flow from the plane of the sky is shown on the legends for each PV diagram.}
\label{pvmod}
\end{figure*}


However, the basic characteristic of a peak velocity profile offset from the peak of the symmetry profile is clearly present when rotation is introduced. Reversing the sense of direction of the rotation also displaces the velocity profile to the other side of the symmetry peak, indicating that it is the sense of rotation that is responsible for this characteristic. This compares quite nicely to what we see in Figure \ref{pvdiag} and lends weight to our suggestion that the HH\,26 and HH\,72 jets are rotating in opposing directions.

The shape of the PV diagram, and the corresponding peak velocity profile, is more sensitive to some model parameters than others. For a given inclination angle, models with steeper power laws for the jet velocity (parameterised by $\gamma$ in equation \ref{vels}) have the effect of introducing a prominent `butterfly-wing' to the PV diagram. However, a weak dependence is seen with how the azimuthal velocity changes with radius, parameterised by $\beta$, if $\gamma$ is fixed. In fact, the radial velocity diagrams remain remarkably similar for all values of $\beta$ although they become more symmetrical as $\beta$ becomes shallower. Any differences between the models are subtle and most easily seen in the velocity diagrams.

In Figure \ref{pvmod} we show models as a function of inclination at fixed jet and azimuthal velocities (see Table \ref{modpars} for the full list of parameters). The emission tracks across the PV diagram as the inclination increases, the jet velocity component becoming more prominent at higher inclinations. We also start to see limb brightened inner edges in the PV diagram, a consequence of the density structure (equation \ref{dens}), with the extent of the `butterfly-wing' shape determined by the velocity power law ($\gamma$). The radial velocity profile becomes more symmetrical at higher inclinations. 

These models show just how instructive PV diagrams can be. Comparison with Figure \ref{pvdiag} suggests that the jets are inclined to the line-of-sight at angles between $30-40\deg$.

No attempt has been made to actually fit a model to the data (the breadth of the parameter space precludes such a search). With the caveats already mentioned, the systematic changes seen in the intensity-weighted radial velocity diagrams are well represented by a jet that rotates about its axis.

\begin{table}
\begin{minipage}[t]{\columnwidth}
\caption{Parameters used in PV diagram model.}             
\label{modpars}      
\centering                       
\renewcommand{\footnoterule}{}  
\begin{tabular}{l c}        
Parameter & Value \\    
\hline
Pixel scale & 0.148$\arcsec$ pixel$^{-1}$ \\
Distance & 400\,pc \\
Vertical Size\footnote{Size of emission as it would appear on the sky.} & 1.0$\arcsec$ \\
Inner radius, $Ri$ & 0.1$\arcsec$ \\
Outer radius, $Ro$ & 2.5$\arcsec$ \\
Seeing (FWHM) & 0.5$\arcsec$ \\
Scale length (density), $a$ & 0.5$\arcsec$ \\
& \\
Jet velocity, $v_{\textrm{jet},0}$ & 150\,\kms \\
Azimuthal velocity, $v_{\phi,0}$ & 3\,\kms \\
Radial velocity, $v_r(r)$ & 0\,\kms \\
Spectral resolution & 33.7\,\kms \\
& \\
Exponent (density), $\alpha$ & 0.75 \\
Exponent (azimuthal velocity), $\beta$ & 0.5 \\
Exponent (jet velocity), $\gamma$ & 0.1 \\
\hline                                   
\end{tabular}
\end{minipage}
\end{table}

\subsection{Analysis of the HH\,26 jet rotation signature}

In the following sections the results for HH\,26 are analysed under the hypotheses that they do indeed indicate the presence of jet rotation, that the mechanism launching the \mh portion of the jet can  be identified as a large scale magneto-hydrodynamic (MHD)  disk wind, (e.g. \cite{Konigl00} and  \cite{Ferreira02}), and the quantities derived in our work can be used to constrain the properties of the jet acceleration region. The derivation follows closely the one described in \cite{Woitas05} for the atomic jet from the classical T Tauri star RW Aur, seen at optical wavelengths. We do not conduct a similar analysis for HH\,72 as there is insufficient information available in the literature to estimate values such as the mass accretion rate and mass flux. Furthermore, its greater distance from us than HH\,26 makes the radial velocity measurements more difficult to interpret. 

In the standard disk-wind model, the flow is assumed to be steady-state, axisymmetric and to
satisfy the ideal MHD equations. In a cylindrical coordinate system ($r, z, \phi$) with the star at the origin, the magnetic field lines identified by the poloidal and toroidal components of the magnetic field  (${\bf B_p } = B_r {\bf e_r} + B_z {\bf e_z} $ and  $B_{\phi} {\bf e_{\phi}} $, respectively) lie on  nested magnetic surfaces with an hourglass shape, intersecting the disk equatorial plane at the so-called `footpoint' (see Fig. 1 in \citet{Woitas05}). The disk and the  anchored field lines rotate 
rapidly, and the centrifugal force  launches fluid parcels from the disk surface out along the oblique open field lines. This in turn generates a ``magnetic torque''  that brakes the disk, and extracts energy and angular momentum from it. In the acceleration process, the matter  reaches a point (the so-called Alfv\'en surface, at a height of a few AUs above the disk) where the inertia of the matter overcomes the magnetic forces, and the field is wrapped tightly generating a very important toroidal component. This in turn produces a magnetic force (``hoop stress'') directed toward the axis, that  collimates  the flow. 

Using our radial velocity measurements, one can attempt to determine the  so-called `footpoint radius' of the wind, that is the location in the disk from where the observed portion of the wind originates, and the ratio between the poloidal and toroidal components of the magnetic field in the observed section of the jet.

First, one has to compute the value of the poloidal and toroidal components of the velocity field, 
${\bf v_p} = v_r {\bf e_r}+ v_z {\bf e_z} $ and $v_{\phi}$ respectively, at the observed location, 
that is at $\sim$ 800\,AU above the disk for HH\,26. These components can be derived from our rotation measurements simply as,
 
\[
v_p \approx v_z  = (v_{\rm rad 1} + v_{\rm rad 2})
/ (2 \cos i)
\]
\begin{equation}
v_{\phi} = (v_{\rm rad 1} - v_{\rm rad 2}) / (2 \sin i ),
\end{equation}

\noindent
where $v_{\rm rad 1,2}$ are the measured radial velocities at two equidistant positions either side of the flow axis (with 1 referring to the  northern part of the slit) and $i = 90\deg - \theta$ is the inclination angle of the flow with respect to the line of sight.  

\begin{figure}
\centering
\includegraphics[width=\columnwidth]{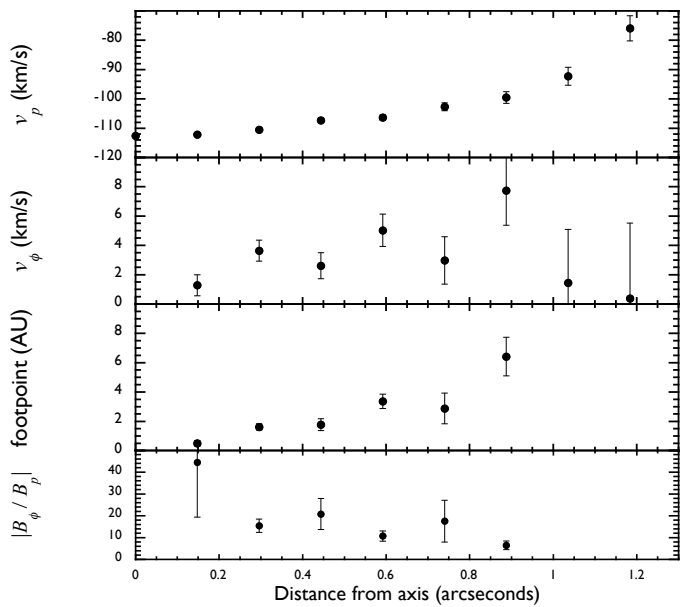}
\caption{Results of the \mh rotation analysis for HH\,26 assuming a flow inclination angle of $40^{\circ}$ with respect to the line of sight and a central mass of 1 M$_\odot$. The values of the various quantities are obtained  from the parameters of the wind observed at the distance from the axis defined by the $x$ coordinate. From top to bottom: poloidal velocity; toroidal velocity derived from the radial velocity shifts; distance from the star of the wind footpoint, according to an ideal MHD disk wind scenario; absolute value of the magnetic pitch angle $|B_{\phi} / B_p|$ at the observed locations in the flow.
}
\label{rotanal}
\end{figure}

The derived values of these quantities are shown in Figure \ref{rotanal}. The errors are propagated in quadrature from the measurement error of the radial velocity. We assumed a value of 40\deg\, for the inclination of HH\,26, that follows from the combination of tangential proper motion measurements ($\sim 70$\,\kms) reported in \cite{Chrysostomou00} for HH\,26 knot C, the closest to our observations, and our measured radial velocity ($-86$ \kms).

The values for $v_p$ and $v_{\phi}$ are of course proportional to those shown in Figure \ref{pvdiag}. The poloidal velocity varies between about $-75$ and $-110$\,\kms while the toroidal velocity values span between 1 and 7 \kms . One notes that while the poloidal velocity shows the expected behaviour, i.e., its magnitude gradually decreasing with distance from the axis, the absolute value of $v_{\phi}$ first increases reaching a plateau before decreasing again, while one would expect that it would have its maximum toward the jet axis. This is due to beam smearing and projection effects of the portions of the jet that are intercepted by the line of sight \citep{Pesenti04}. This effect may introduce an underestimate of the toroidal velocity by up to 20\% in the two positions closest to the axis, which also affects the determination of the footpoint radius and of the magnetic pitch angle at the same position, as explained below. On the other hand one can see that the regular trend in  $v_p$ and $v_{\phi}$ is interrupted in the last 2-3 outermost positions, where the errors are largest. This indicates either an objective difficulty in determining the velocity, and/or the fact  that the borders of the knot are interacting strongly with the surrounding environment, and the material emitting in these positions is not participating to the general motion of the flow. For this reason the points at  $1-1.2$\arcsec\, from the axis are excluded from the subsequent analysis. In conclusion, the most reliable estimates are obtained from data at positions $0.3 - 0.6$\arcsec\, from the axis.

\subsubsection{Footpoint radius}

The third panel down in Figure \ref{rotanal} reports our estimate of the footpoint radius $r_{0}$ for the wind, i.e. the region in the disk from where the matter detected in our data at a distance $r_{\mathrm{obs}}$ from the flow axis should originate. This can be derived from the combination of poloidal and toroidal velocities measured at $r_{\mathrm{obs}}$. In fact, in a magneto-centrifugal wind, the footpoint radius of the jet component located at $r_{\rm obs}$  from the axis with measured $v_{\phi}$ and $v_p$ can be estimated using a relationship provided by \cite{Anderson03} which is valid at large distances from the source:

\[
r_0 \approx  0.7\,\mathrm{AU} \left( {r_{\mathrm{obs}} \over 10\,\mathrm{AU}}  \right)^{2/3}   
\left( {v_\phi \over 10\,\mathrm{km}\,\mathrm{s}^{-1} }  \right)^{2/3}
\]
\begin{equation}
~~~~~~~~~~~~~~~~~~~~~~~    
\times
\left( {v_{\mathrm{p}}  \over 100\,\mathrm{km}\,\mathrm{s}^{-1}} \right)^{-4/3}   
\left( {M_\star \over 1~~ M_\odot}  \right)^{1/3}.
\label{errezero}   
\end{equation} 

This simple equation is valid under the condition $v_{\phi} \ll v_{\mathrm{p}}$, which in the context of disk winds is achieved as soon as the Alfv\'en surface is reached, at a few AUs above the disk. Thus at the sampled distances from the star, one can be confident that this asymptotic regime has been reached by the outer streamlines of the flow. Although we don't precisely know the mass of the central protostar, with a luminosity of $\sim 5$\,L$_\odot$ \citep{Antoniucci07} and considering that only a small fraction of this luminosity is due to accretion one can infer that HH\,26 has a mass of $\sim 1$\,M$_\odot$. 
Once again, the errors are propagated quadratically from the measured  quantities. 

Using equation (\ref{errezero}) we obtain footpoint radii varying between 0.5 to 6 AU from the protostar.  Keeping in mind the caveat for measured velocities close to and far from the flow axis, one should take the extreme values with caution and accept that a more reliable estimate of the footpoint radius for this jet lies between 2 and 4 AU for a star of 1 M$_\odot$\, (with an average error of 20\%). 

These values are consistent with those of \cite{Coffey04} and \cite{Woitas05} when compared to the footpoints of jets emitting in optical lines. For those, which sample a portion of the flow nearer the axis, footpoint radii between 0.5 and 2-3 AU were found. This is in agreement with the notion that the molecular wind should be produced in a region of the disk with low ionisation external to the one from which optical jets originate, as found for DG Tau \citep{Bacciotti00,Takami04}.  

It should be noted that the above calculation gives only the footpoint of the flow surface for which the toroidal velocity can be measured in the jet, and {\em not} the radius of the whole ejection region. In the disk-wind scenario, other colder components seen in molecular lines at longer wavelengths may be anchored to larger footpoint radii.

\subsubsection{Magnetic field properties}

In the hypothesis of disk-wind acceleration for the jet, our measurements allow one to easily find the ratio  between the toroidal and poloidal components of the magnetic field, $B_{\phi} / B_p$, in the jet  at the observed  location. This quantity indicates by how much the lines of force are wrapped on a given magnetic surface. The MHD models for jet acceleration prescribe that the collimation arises from the hoop stress generated by the increase of the toroidal component of the magnetic field above the Alfv\`en point (see, e.g. \citet{Konigl00}). It is thus interesting to check if the observations give any indication for the magnetic field configuration consistent with this property in the region of the flow just above the collimation zone. The ratio  $B_{\phi} / B_p$ is found using another conservation law of general disk-wind theory \citep{Konigl00,Anderson03}:  

\begin{equation}
\left(v_{\phi} - {B_{\phi} v_p \over B_p} \right) / r  = \Omega_0,  
\label{eq:pitch}
\end{equation} 
\noindent
where $\Omega_0$ is the disk angular velocity at the footpoint. 

Introducing our measured velocities at $r= r_{\mathrm{obs}}$ and using the corresponding $r_0$ to calculate $\Omega_0$ for a Keplerian disk, we find  $B_{\phi}/ B_p$ for any $r= r_{\mathrm{obs}}$. The results are shown in the bottom panel of Figure \ref{rotanal} in absolute terms (the sign of the ratio is consistent with that of the velocity components and the assumed geometry). Once again caution should be taken for points both close and far from the axis. Good estimates are probably given by the three positions between $0.3-0.6$\arcsec, for which we find $|B_{\phi}/ B_p|$ varying from 10 to 20.

Thus, the toroidal component of the magnetic field appears to be dominant, which supports the suggestion that  protostellar outflows are collimated magnetically, as prescribed by the MHD acceleration models.

\subsubsection{Angular momentum transport}

In principle, it is possible to use our measurements to estimate the amount of angular momentum carried by the jet. Following magneto-hydrodynamical models, the angular momentum transported along each flow surface of the jet is a constant along the surface and has both a kinetic and magnetic component that always add in sign. The total angular momentum transported by the flow can then be calculated by integrating the contributions of all the flow surfaces across the jet. For the two systems for which this has already been done (namely, DG\,Tau and RW\,Aur), the angular momentum carried away was estimated to be between 60\% and 100\% of the angular momentum that the inner disk has to lose to accrete at the observed rate \citep{Bacciotti02,Woitas05}. The fundamental implication of this is that jets are likely to be the major agent for extracting excess angular momentum from the star/disk system which allows matter to flow through the disk and accrete onto the star.

We cannot undertake a similarly accurate calculation for HH\,26 as the internal distribution of the molecular gas in the jet, and therefore the location of the most internal flow surface of the \mh-emitting gas, is not precisely known. This affects the calculation of the integral of the angular momentum and in particular the determination of the magnetic contribution \cite[see][]{Woitas05}. Nevertheless, it is  possible to approximate the {\em kinetic} contribution, a lower limit to the actual angular momentum transported, and compare this part with the angular momentum that has to be lost by the disk. The kinetic angular momentum can be expressed as,
  
\begin{equation}
\dot{L}_{\mathrm{jet}} ~ = ~ \int_{S} \left( r v_{\mathrm{\phi}}\right) \rho {\bf v_{\mathrm{p}}} \cdot {\bf n}~ dS \approx \bar{( r v_{\mathrm{\phi}})}\dot{M}_{\rm jet},
\label{eq:angmom} 
\end{equation} 

\noindent
where $S$ is the surface of the observed section of the jet, $\rho$ is the total mass density,  $\bar{( r v_{\mathrm{\phi}})}$ is the average value of the specific angular momentum in the observed region and  $\dot{M}_{\rm jet}$ is the jet mass flux in \mh.

For HH\,26, the \mh mass flux rate in the internal knot considered here has been computed in \cite{Antoniucci07} who derive a value between $2-5\times10^{-8}$\,M$_{\odot}$ yr$^{-1}$. The same authors also estimate a mass accretion rate of $8.5 \times 10^{-7}$\,M$_{\odot}$ yr$^{-1}$ from the luminosity of the Br$\gamma$ line. Assuming that the toroidal velocity measured at $r_{\rm obs} = 0.44$\arcsec\,, $v_\phi=2.5$\,\kms is representative of the average over the \mh emitting region, one obtains for a distance of 400 pc, $\dot{L}_{\mathrm{jet}} \sim  2\times10^{-5}$  M$_\odot$ yr$^{-1}$ AU km s$^{-1}$. For comparison the total angular momentum transported by the optical jet in the RW Aur system was estimated to be $\sim 2.5 \times10^{-5}$  M$_\odot$ yr$^{-1}$ AU km s$^{-1}$ in each lobe.

It is now interesting to compare this with an estimate of the angular momentum the disk has to lose for the star to accrete at the observed rate. The excess disk angular momentum can be found through the expression \citep[cf.][]{Woitas05}:

\begin{equation}
\label{excess}
\dot{L}_{\mathrm{disk}} = (\dot{M}_{\mathrm{acc}} + \dot{M}_{\mathrm{flow}}) ~ r_{0,{\mathrm{ext}}}~ v_{\mathrm{K,ext}} - \dot{M}_{\mathrm{acc}}~ r_{0,\mathrm{in}}~ v_{\mathrm{K,in}}
\end{equation}

\noindent where $\dot{M}_{\mathrm{acc}}$ is the mass accretion rate onto the star, $\dot{M}_{\mathrm{flow}}$ is the sum of the mass loss rates in the two lobes of the whole flow (atomic and molecular components), $r_{0,{\mathrm{in,ext}}}$ are the inner- and outermost footpoint radii of the whole flow and $v_{\mathrm{K,in,ext}}$ are the corresponding Keplerian velocities. Both magneto-hydrodynamic models and observations prescribe that $\dot{M}_{\mathrm{flow}} = 0.1\, \dot{M}_{\mathrm{acc}}$, and that $r_{0,\mathrm{in}} \sim 0.03$\,AU from the star. 

Using our estimate of the outermost footpoint of 6 AU from the star together with the mass accretion rate estimated by \cite{Antoniucci07} we obtain, for a 1 solar mass star, $\dot{L}_{\mathrm{disk}} \sim 6 \times 10^{-5}$\,M$_\odot$\,yr$^{-1}$\,AU\,\kms. Assuming that HH\,26 has a counterjet that possesses similar physical properties, we find that the kinetic angular momentum transported by the bipolar \mh jet already amounts to about 70\% of the angular momentum that has to be extracted from the disk. Considering that the kinetic component of the molecular jet gives just a lower limit to the actual total angular momentum in the flow, our result lends further support to the hypothesis that jets are the main agent for extraction of excess angular momentum in the disk.

\subsection{Caveats and alternative explanations}

The results and analysis presented here seem to indicate that, for HH\,26 at least, the observed portion of the jet originates from an extended region of the circumstellar disk. We cannot exclude, however, the presence of an inner X-wind or stellar wind (cf. \cite{Shu00} and \cite{Sauty03}), as at the resolution of our observations we only probe the external flow surfaces. In addition, it is not yet proven that flow lines can be entirely traced back from the observed jet locations to the footpoints in a real jet. In fact, once the region beyond the acceleration zone is reached, various magneto-fluid instabilities can complicate the geometry of the field lines and a direct connection to the footpoints may be lost. 

Furthermore, a number of recent studies have proposed alternate explanations for the observed velocity asymmetries. In fact, there are aspects that still await clarification, the most intriguing of which to date is the recent study of the kinematics of the RW Aur disk that suggest a rotation sense for the disk opposite to the one of the bipolar jet \citep{Cabrit06}. At the moment, therefore, alternative explanations for the observational evidence are actively searched for. According to recent studies, velocity asymmetries may also be produced by interaction with a warped disk \citep{Soker05}, or in asymmetric shocks generated by jet precession \citep{Cerqueira06}. However, in their simulations of rotating and/or precessing jets, \citet{Smith07} find no evidence of rotation signatures being mimicked by a precessing jet, although they do warn that any rotation that is present will quickly be dissipated as the jet expands. Interestingly, they predict that jet rotation introduces instabilities in the flow which develop into a ``swarm of bow features", rather like those observed ahead of our slit placements (see Fig. \ref{slits}). Such interpretations need to be able to explain the statistics (in practice 100\% of examined cases) with which velocity asymmetries are found, and within the appropriate range predicted for rotation. 


\section{Conclusions}

We present observations of the \mh emission from jets of two embedded Class I sources, HH\,26 and HH\,72, and search for velocity asymmetries in the spectral lines from \textit{across} the jets that are compatible with rotation. Our results are consistent with those signatures found at optical wavelengths for T Tauri jets, but for objects at an earlier evolutionary phase.

The position-velocity diagrams for HH\,26 and HH\,72 show a tilt in the intensity distribution of the PV diagrams, i.e. an asymmetry of the jet velocities with respect to the axis which can be interpreted as rotation of the gas. The observed tilt is opposite in the two jets indicating that the sense of rotation is different for the two flows.

The rotation hypothesis is tested by constructing synthetic PV diagrams using an analytic model for the \mh radiation emitted by a cylindrical jet observed by a spectrograph slit. The observed skew in the PV diagrams can be reproduced assuming that the jet material possesses a toroidal velocity of the order of a few \kms which decreases as a simple power-law with the jet radius, together with a jet velocity that also decreases with distance from the jet axis.

Under the assumption that the flow is indeed rotating, and that it is launched magneto-centrifugally, one can apply very general conservation equations valid for such jet acceleration mechanisms to retrieve information on the physics of the jet itself. One can show that the part of the flow under consideration may originate in an annulus of the disk located at a distance from the star of 2-4\,AU for HH\,26. This agrees with the notion that the molecular wind should be produced in a region of the disk external to the one from which the optical jets originate \citep{Takami04} --- up to 2--3\,AU in the disk according to previous rotation studies. In a similar way it is possible to show that the toroidal component of the magnetic field in the observed portion of the flow is dominant over the poloidal component, in analogy with what has been found for optical jets and in support of a magnetic collimation mechanism.

On the basis of our measurements and previous determination of mass loss rates, we estimate that the kinetic angular momentum transported by the HH\,26 jet is about $2\times10^{-5}$\,M$_{\odot}$\,yr$^{-1}$\,AU\kms. This is a lower limit to the total angular momentum transported, which also includes a magnetic contribution (not derivable from our measurements), but already amounts to 70\% of the angular momentum that has to be lost by the disk for the star to accrete at the observed rate. A similar estimate for HH72 has not been possible, due to uncertainties introduced by its large distance and the lack of an estimate of the accretion rate onto the source.

These results are in agreement with most model prescriptions of protostar formation according to which the jets are produced in the system as soon as the disk begins to accrete material onto the central protostar. This occurs because of the need to remove excess angular momentum from the central accreting system. Further observational studies, using different tracers and conducted on a larger number of jets as well as their associated disks, need to be made towards objects at various evolutionary phases in order to prove whether rotation is a general property of stellar outflows.


\begin{acknowledgements}
Based on observations collected at the European Southern Observatory, Chile (ESO Programme 72.C-0054). This work was partially supported by the European Community's Marie Curie Research \& Training Network ``Jet Simulations, Experiments and Theory (JETSET)", under contract MRTN-CT-2004-005592.
\end{acknowledgements}



\begin{thebibliography}{}

\bibitem[Anderson et al.(2003)]{Anderson03} Anderson, J.M., Li, Z.-Y., Krasnopolsky, R., Blandford, R., 2003, ApJ, 590, L107

\bibitem[Anthony-Twarog(1982)]{Anthony-Twarog82} Anthony-Twarog, B.J., 1982, AJ, 87, 1213

\bibitem[Antoniucci et al.(2008)]{Antoniucci07} Antoniucci, S., Nisini, B., Gianni, T., Lorenzetti, D., 2008, A\&A, in press. [astro-ph/0710.5609]

\bibitem[Bacciotti et al.(2000)]{Bacciotti00}	Bacciotti, F., Mundt, R., Ray, T.P., Eisl\"{o}ffel, J., Solf, J., Camenzind, M., 2000, ApJ, 537, 49

\bibitem[Bacciotti et al.(2002)]{Bacciotti02} Bacciotti, F., Ray, T.P., Eisl\"{o}ffel, J., Solf, J., 2002, ApJ, 576, 222

\bibitem[Bally, Reipurth \& Davis(2007)]{Bally07} Bally, J.,  Reipurth, B., \& Davis, C.J., 2007, in \textit{Protostars \& Planets V}, B. Reipurth, D. Jewitt, K. Keil eds., University of Arizona Press, Tuscon, p.215

\bibitem[Black \& van Dishoeck(1987)]{BvD87} Black, J.H., \& van Dishoeck, E.F., 1987, ApJ, 322, 412

\bibitem[Cabrit et al.(2006)]{Cabrit06} Cabrit, S., Pety, J., Pesenti, N., Dougados, C., 2006, A\&A, 452, 897

\bibitem[Calzoletti et al.(2008)]{Calzoletti08} Calzoletti, L., Giannini, T., Nisini, B., et al. 2008, in preparation

\bibitem[Cerqueira et al.(2006)]{Cerqueira06} Cerqueira, A.H., Velazquez, P.F., Raga, A.C., Vasconcelos, M.J., de Colle, F., 2006, A\&A, 448, 231

\bibitem[Chrysostomou et al.(2000)]{Chrysostomou00} Chrysostomou, A., Hobson, J., Davis, C.J., Smith, M.D., Berndsen, A., 2000, MNRAS, 314, 229

\bibitem[Coffey et al.(2004)]{Coffey04} Coffey, D., Bacciotti, F., Woitas, J., Ray, T.P., \& Eisl\"{o}ffel, 2004, ApJ, 604, 758

\bibitem[Coffey et al.(2007)]{Coffey06} Coffey, D., Bacciotti, F., Ray, T.P., Eisl\"{o}ffel, J., Woitas, J., 2007, ApJ, 663, 350

\bibitem[Davis et al.(2000)]{Davis00} Davis, C.J., Berndsen, A., Smith, M.D., Chrysostomou, A., Hobson, J., 2000, MNRAS, 314, 241

\bibitem[Davis et al.(2001)]{Davis01} Davis, C.J., Ray, T.P., Desroches, L., Aspin, C., 2001, MNRAS, 326, 524

\bibitem[Davis et al.(2002)]{Davis02} Davis, C.J., Stern, L., Ray, T.P., Chrysostomou, A., 2002, A\&A, 382, 1021

\bibitem[Ferreira (2002)]{Ferreira02} Ferreira, J., 2002, EAS Publications Series, Volume 3, Proceedings of "Star Formation and the Physics of Young Stars", held 18-22 September, 2000 in Aussois France. Edited by J. Bouvier and J.-P. Zahn. EDP Sciences, 2002, p. 229

\bibitem[Ferreira et al.(2006)]{Ferreira06} Ferreira, J., Doiugados, C., \& Cabrit, S., 2006, A\&A, 453, 785

\bibitem[K\"onigl \& Pudritz(2000)]{Konigl00} K\"{o}nigl, A., \& Pudritz, R.E., 2000, in Protostars and Planets IV, Mannings, V., Boss, A.P., Russell, S. S., eds., University of Arizona Press, Tuscon, p. 759

\bibitem[Pesenti et al.(2004)]{Pesenti04} Pesenti, N., Dougados, C., Cabrit, S., Ferreira, J., Casse, F., Garcia, P., \& O'Brien, D., 2004, A\&A, 416, L9

\bibitem[Podio et al.(2006)]{Podio06} Podio, L., Bacciotti, F., Nisini, B., Eisl{\"o}ffel, J., Massi, F., Giannini, T., \& Ray, T.P., 2006, A\&A, 456, 189

\bibitem[Pudritz et al.(2007)]{Pudritz06} Pudritz, R., Ouyed, R., Fendt, C., Brandenburg, A., 2007, in \textit{Protostars \& Planets V}, B. Reipurth, D. Jewitt, K. Keil eds., University of Arizona Press, Tuscon, p.277

\bibitem[Ray (2000)]{Ray00} Ray, T.P., 2000, ApSS, 272, 115

\bibitem[Ray et al.(2007)]{Ray06} Ray, T.P., Bacciotti, F., Dougados, C., Eisl\"{o}ffel, J., Chrysostomou, A., 2007, in \textit{Protostars \& Planets V}, B. Reipurth, D. Jewitt, K. Keil eds., University of Arizona Press, Tuscon, p.231

\bibitem[Rousselot et al.(2000)]{ohlines} Rousselot, P., Lidman, C., Cuby, J.-G., Moreels, G., Monnet, G., 2000, A\&A, 354, 1134

\bibitem[Sauty et al.(2003)]{Sauty03} Sauty, C., Tsinganos, K., Trussoni, E., Meliani, Z., 2003, Astrophysics and Space Science, 287, 25

\bibitem[Shang et al.(2007)]{Shang06} Shang, H., Li, Z.-Y., Hirano, N., 2007, in \textit{Protostars \& Planets V}, B. Reipurth, D. Jewitt, K. Keil eds., University of Arizona Press, Tuscon, p.261

\bibitem[Shu et al.(2000)]{Shu00} Shu, F. H., Najita, J. R., Shang, H., Li, Z.-Y., in Protostars and Planets IV, Mannings, V., Boss, A.P., Russell, S. S., eds., University of Arizona Press, Tuscon, p. 789

\bibitem[Smith \& Rosen(2007)]{Smith07} Smith, M.D., \& Rosen, A., 2007, MNRAS, 378, 691

\bibitem[Soker(2005)]{Soker05} Soker, N., 2005, A\&A, 435, 125

\bibitem[Takami et al.(2004)]{Takami04} Takami, M., Chrysostomou, A., Ray, T. P., Davis, C., Dent, W. R. F., Bailey, J., Tamura, M., Terada, H., 2004, A\&A, 416, 213

\bibitem[Woitas et al.(2005)]{Woitas05} Woitas, J., Bacciotti, F., Marconi, A., Coffey, D., Eisl\"{o}ffel, J., 2005, A\&A, 432, 149


\end{thebibliography}
\end{document}